\documentclass[twocolumn,showpacs,preprintnumbers,amsmath,amssymb]{revtex4}

\usepackage{graphicx}% Include figure files
\usepackage{graphicx,epstopdf}
\usepackage{xcolor}
\usepackage{dcolumn}% Align table columns on decimal point
\usepackage{bm}% bold math
\usepackage{mathrsfs} % script-like, curvy letters.

\begin{document}

%\preprint{APS/123-QED}
\title{Chimera patterns in three-dimensional locally coupled systems}
\author{Srilena Kundu$^1$}
\author{Bidesh K. Bera$^{1,2}$}
\author{Dibakar Ghosh$^1$}
\email{diba.ghosh@gmail.com}
\author{M. Lakshmanan$^3$}
\affiliation{$^1$Physics and Applied Mathematics Unit, Indian Statistical Institute, 203 B. T. Road, Kolkata-700108, India\\
	$^2$Department of Mathematics, Indian Institute of Technology Ropar, Punjab-140001, India\\
	$^3$Centre for Nonlinear Dynamics, School of Physics, Bharathidasan University, Tiruchirapalli-620024, India}

\date{\today}

\begin{abstract}
The coexistence of coherent and incoherent domains, namely the appearance of chimera states, is being studied extensively in many contexts of science and technology since the past decade, though the previous studies are mostly built on the framework of one-dimensional and two-dimensional interaction topologies. Recently, the emergence of such fascinating phenomena has been studied in three-dimensional (3D) grid formation while considering only the nonlocal interaction. Here we study the emergence and existence of chimera patterns in a three dimensional network of coupled Stuart-Landau limit cycle oscillators and Hindmarsh-Rose neuronal oscillators with local (nearest neighbour) interaction topology. The emergence of different types of spatiotemporal chimera patterns is investigated by taking two distinct nonlinear interaction functions. We provide appropriate analytical explanations in the 3D grid of network formation and the corresponding numerical justifications are given. We extend our analysis on the basis of Ott-Antonsen reduction approach in the case of Stuart-Landau oscillators containing infinite number of oscillators. Particularly, in Hindmarsh-Rose neuronal network the existence of non-stationary chimera states are characterized by instantaneous strength of incoherence and instantaneous local order parameter. Besides, the condition for achieving exact neuronal synchrony is obtained analytically through a linear stability analysis. The different types of collective dynamics together with chimera states are mapped over a wide range of various parameter spaces.  
\end{abstract}


\maketitle

%\tableofcontents

\section{Introduction}\label{intro}
The study on the collective behaviour of coupled dynamical systems has drawn a significant attention during the past two decades due to its immense applicability in several physical, chemical, biological and engineering systems. Among the varied collective dynamics, chimera state \cite{chimera_rev,epl_rev,plr} is one of the most studied research topics in recent times and since its identification, the collective dynamics of identical oscillators have been revitalized. Chimera state \cite{kuramoto,strogatz} is a complex spatiotemporal pattern that consists of both coherent and incoherent domains simultaneously and appear in a network of symmetrically coupled identical oscillators. The existence and robustness of chimera states has been verified through a series of experiments in different systems, including chemical oscillators \cite{chem_chi}, electrochemical oscillators \cite{elchem_chi}, electronic circuits \cite{elec_chi}, opto-elctronic systems \cite{opto_chi}, mechanical systems \cite{mech_chi} and frequency modulated time delay systems \cite{feddback_chi}. 

\par Some aquatic mammals such as eared seal, dolphin, some migratory birds and manatees are engaged in the slow wave unihemispheric sleep \cite{uhsws1,uhsws2}. During their sleeping state, half of their brain is shut down while the remaining portion of the hemisphere is awake to monitor what is happening in the environment. The neuronal oscillations in the wake portion of the cerebral hemisphere is desynchronized whereas oscillations in the sleepy part is very much synchronized. This type of neuronal process in the brain is closely related to Kuramoto's observation of the coexistence of coherent and incoherent motions in a network of nonlocally coupled identical phase oscillators \cite{kuramoto}. Later, Abrams {\it et al.} \cite{strogatz} provided some theoretical explanations for the existence of such states and named it as {\it chimera state}. The chimera-like features are also observed in many man-made systems and natural phenomena such as power grid networks \cite{power1,power2}, superconducting metamaterial \cite{scm}, ventricular fibrillation \cite{vf1,vf2} and different types of brain diseases \cite{brain_disease1,brain_disease2}, like epileptic seizures, brain tumors and schizophrenia. The chimera state is a fascinating type of symmetry-breaking complex pattern and appears in symmetry preserving systems. After the detection of chimera states in a  network of phase oscillators \cite{kuramoto,strogatz}, the existence of these states has also been proved in networks of chaotic \cite{chaotic} and hyper chaotic oscillators \cite{lakshman_measure}, limit cycle oscillators \cite{limit,limit2}, chaotic maps \cite{chaotic_map}, time delayed oscillators \cite{bs_chimera} and even in neuronal oscillators \cite{hr_ijbc} which exhibit different types of bursting dynamics. At the beginning it was believed that nonlocal network configuration is the essential requirement for the existence and emergence of chimera states since such coupling topology produces multistability in the systems and leads to the appearance of such a state. But recent results revealed that proper design of the coupling functions and the nodal dynamics may generate such states in globally (all-to-all) \cite{global1,global2,global3,global4,global5,global6,global7} coupled networks and locally (nearest neighbour)  coupled rings \cite{laing,hr_bera1,hr_bera2,local1} and arrays \cite{limit2} of oscillators. Besides these symmetric interacting configurations, the existence of chimera states has also been uncovered in random and scale free networks \cite{complex_ch1}, heterogeneous \cite{ch_hetero}, multiplex networks \cite{chimera_multiplex,multiplex1,multiplex2,multiplex3,multiplex4}, time varying networks \cite{complex_ch2}, modular networks \cite{chimera_modular} and  multiscale networks \cite{chimera_multiscale}. The existence of chimera states is also detected in two interacting different populations where each oscillator in each population interacts in nonlocal \cite{two_nonlocal} and global fashion \cite{two_global}.  Recently, the emergence of alternating chimera in a network of identical neuronal systems induced by an external electromagnetic field has been reported in \cite{chimera_ephaptic}. Depending on the spatiotemporal motion of the oscillators in the network, different types of non-stationary chimera patterns are categorized which include imperfect chimera \cite{imperfect_chi}, travelling chimera \cite{travelling_chi}, imperfect travelling chimera \cite{hr_bera3}, and spiral wave chimera states \cite{spiral_chi,epjst_spiral}. From the amplitude variation of each individual oscillators of the network, various types of chimera states are classified such as chimera death \cite{cd_prl}, amplitude mediated chimera \cite{amc} and so on. 
          
\par In real world systems, the human brain is one of the most challenging complex systems due to the presence of large number of neurons and myriads of interconnections between them. Also the understanding of the neuronal communication and interaction patterns between the neurons through the synapses is also a real challenging issue. Mainly two structurally different types of synapses are identified in the interneuronal communications, one is electrical gap junction and another is chemical synapse. In the electrical communication, the neuronal information are exchanged between two adjacent neurons by making the gap junction channel in the membrane potential, while through the chemical synapses the signal is transferred chemically in the form of neurotransmitter molecules among the neurons. Normally, neurons in the brain are situated and pass information in a two dimensional (2D) grid formation for their normal operation. The existence and emergence of the chimera states in the 2D grid of oscillations has been investigated in the neuronal network by taking electrical and chemical synapses in nonlocal \cite{2d1} and local \cite{2d2} modes.  Also, recently, the presence of chimera patterns have been reported in 3D network of neuron oscillators having nonlocal interaction topology \cite{3dhizanidis}.

\par In this work, we systematically investigate the chimera patterns in networks of locally coupled oscillators in 3D grid formation, which can be considered as a generalization of the 2D grid network.  Here we consider the oscillators in 3D grid to be interacting via a nonlinear coupling function along with local coupling topology. Previously chimera state was observed to appear in 1D \cite{hr_bera1} and 2D \cite{2d2} grid of locally coupled oscillators with nonlinear interaction function. Here also we observe that the presence of nonlinearity in the coupling function plays a crucial role for the existence of chimera states in 3D grid of locally coupled oscillators. To understand the mechanism of such emerging phenomena, we start with a  network of locally coupled Stuart-Landau oscillators in the 3D grid formation where the interaction takes place through a nonlinear coupling function. First we numerically show the existence and emergence of chimera states, whose stationarity is justified by the long term spatiotemporal plot, and then by using Ott-Antonsen (OA) reduction method, we perform relevant calculations on the continuum limit approach. The results obtained from this approach are found to be in good agreement with the numerical findings. Next we extend our study on the realistic neuronal network where the Hindmarsh-Rose neurons are considered as local dynamical units of each node of the network. Using nonlinear chemical synaptic interaction function in the local coupling topology, the emergence of chimera states are articulated. In this case, our observed chimera pattern is of a non-stationary type which means that the coherent and incoherent populations of the chimera states are changing erratically in space and time. The existence of such spatiotemporal dynamics is characterized through instantaneous strength of incoherence and local order parameter. Moreover, the analytical bound for exact synchronization in the case of synaptically coupled neuronal network in 3D local interaction topology is derived using a linear stability analysis, and further verified by corresponding numerical results using the notion of Kuramoto order parameter. We further explore the transitions between the several dynamical states and map them over a wide range of different parameter spaces.    

\par The remaining parts of the paper are organized as follows. In Sec. \ref{gmf}, we present a general mathematical description of the three dimensional grid of oscillators. The numerical and analytical investigations on the existence and emergence of chimera patterns in a 3D grid of Stuart-Landau oscillators is provided in Sec. \ref{sl}. Then the subsequent Sec. \ref{hrg} focusses on the occurrence of chimera patterns in the 3D grid of neuronal network and in the sub Sec. \ref{st}, we derived the analytical condition for the neuronal synchrony threshold. The conclusions of our findings are presented in Sec. \ref{con}. 

\begin{figure}[ht]
	\centerline{
		\includegraphics[scale=0.6]{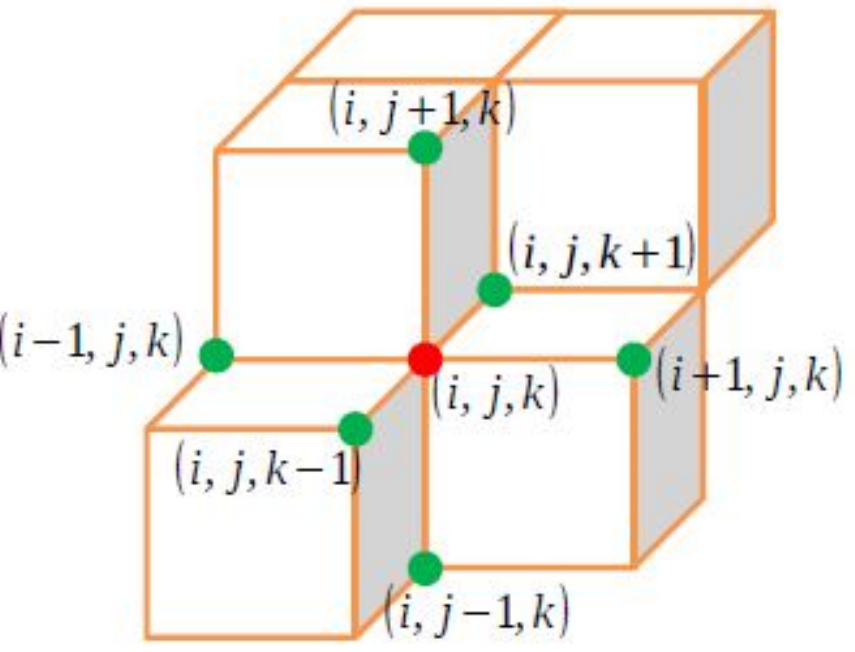}}
	\caption{Schematic diagram of the coupling scheme in 3D cubic lattice: the $(i,j,k)$-th oscillator (red circle) is connected to its six nearest-neighbour oscillators (green circles).}
	\label{fig1}
\end{figure}

	\begin{figure*}[ht]
		\centerline{
			\includegraphics[scale=0.55]{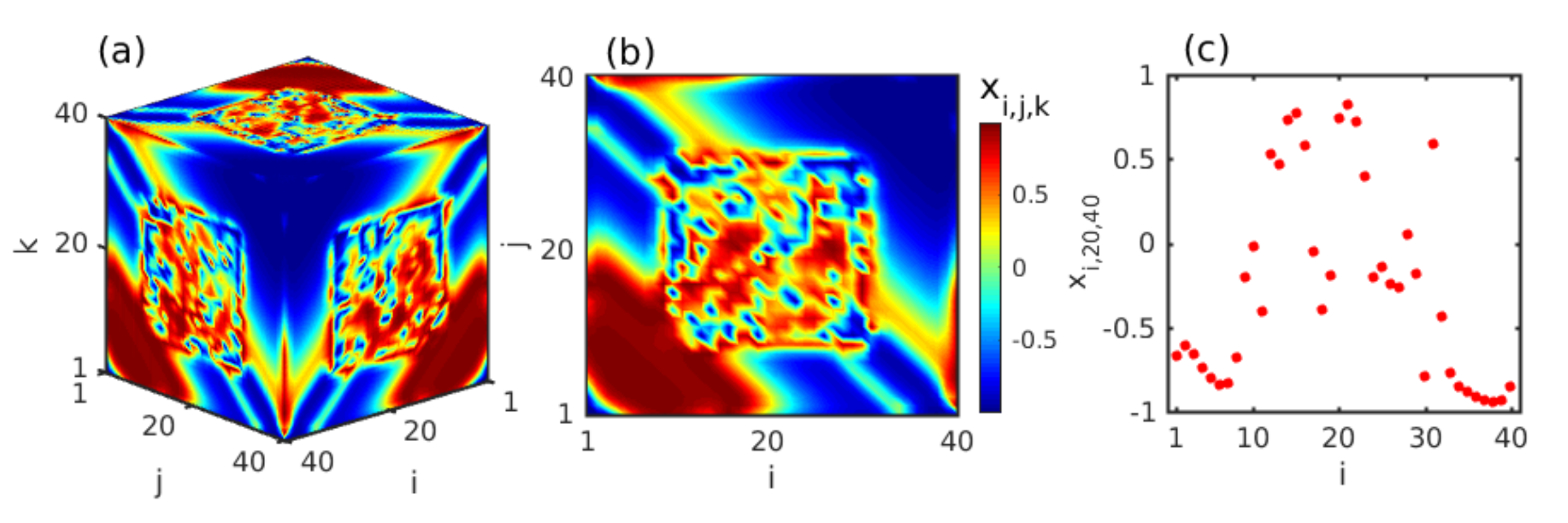}}
		\caption{ Snapshots of the state variables $x$(real part of $z$) in the (a) 3D grid of $N^3$ oscillators, (b) 2D plane with $N^2$ oscillators keeping $k=40$ fixed, (c) in 1D with $N$ oscillators keeping $j=20, k=40$ fixed, showing the emergence of chimera states in the whole network of Stuart-Landau oscillators \eqref{eq:1}. Here $\alpha=1.0, \beta=-1.5, \tilde a=1.02, \epsilon=0.09$, and  $N=40$ are considered.}
		\label{isrevisefig1}
	\end{figure*}
	\section{General mathematical form of 3D coupled oscillators}\label{gmf}
	
	We here consider a network of $ N\times N \times N$ cubic lattice of locally coupled oscillators. The systematic interaction scheme is shown in Fig. \ref{fig1}. The local dynamics of an individual node of the network is given by $\dot X_{i,j,k}=F(X_{i,j,k}),$ where $X_{i,j,k}$ represents an $l$-dimensional vector of the dynamical state variables and $F(X_{i,j,k})$ is the corresponding velocity field. The general mathematical equations of locally coupled systems in a 3D cubic lattice of network can be described as
	\begin{widetext}
		\begin{equation}\label{gm}
		\begin{array}{lcl}
		\dot X_{i,j,k}=F(X_{i,j,k})+K\{ H(X_{i,j,k},X_{i-1,j,k})+H(X_{i,j,k},X_{i+1,j,k})+H(X_{i,j,k},X_{i,j-1,k})+H(X_{i,j,k},X_{i,j+1,k})\\~~~~~~~~~~+H(X_{i,j,k},X_{i,j,k-1})+H(X_{i,j,k},X_{i,j,k+1})\},
		\end{array}
		\end{equation}
	\end{widetext}
	where the subscripts $(i,j,k)(i,j,k=1,2,...,N)$   in $X_{i,j,k}$ and $F(X_{i,j,k})$ determine the position of the oscillator in the 3D coupled network. The coupling function $H : R^l\times R^l\rightarrow R$ describes the manner by which the $(i,j,k)$-th oscillator is connected with its nearest neighbouring oscillators. Here $K=(\epsilon_1,\epsilon_2,...,\epsilon_l)^T$ is the coupling matrix, where $T$ denotes transpose of a matrix. We use periodic boundary conditions as $X_{0,j,k}=X_{N,j,k}$, $X_{N+1,j,k}=X_{1,j,k}$ and $X_{i,0,k}=X_{i,N,k}$, $X_{i,N+1,k}=X_{j,1,k}$ and $X_{i,j,0}=X_{i,j,N}$, $X_{i,j,N+1}=X_{i,j,1}$.
	
	\par In the subsequent sections, we provide adequate evidence for the existence and emergence of the chimera states in 3D lattice of locally coupled oscillators by considering two different dynamical systems, namely Stuart-Landau oscillators and Hindmarsh-Rose neuronal model. For Stuart-Landau oscillators, we consider ``pull-push" type nonlinear coupling function and chemical synaptic interaction function is used in the Hindmarsh-Rose neuronal network.	
	
	\section{Coupled Stuart-Landau system}\label{sl}

	We start our investigation by considering a network of coupled identical Stuart-Landau (SL) oscillators in the three dimensional grid formation where each node of the network is interacting with the local neighbouring nodes. The dynamical evolution of the prescribed network is described by following set of coupled equations, 
		
		\begin{equation} \label{eq:1}
			\begin{array}{lcl}
			\dot{z}_{i,j,k} = (1+\hat{i}\alpha)z_{i,j,k} - (1+\hat{i}\beta)|z_{i,j,k}|^2z_{i,j,k} +\\\\ \frac{\epsilon}{6}[H(z_{i-1,j,k})+H(z_{i+1,j,k}) + H(z_{i,j-1,k}) + H(z_{i,j+1,k}+ \\\\H(z_{i,j,k-1})+ H(z_{i,j,k+1}) - 6H(z_{i,j,k})],
			\end{array}
			\end{equation}
			for $i,j,k=1,...,N$ and $N$ is the total number of oscillators in each direction of the three grid network. The variable $z = x + \hat{i}y$ is a complex variable with $\hat{i}=\sqrt{-1}$ and $\alpha$, $\beta$ are real parameters. The parameter $\epsilon$ denotes the coupling strength and $H(z) = \tilde{a}^2z - z|z|^2,$ represents the nonlinear interaction function \cite{moving_pre} with real constant $\tilde{a}$.  In contrast, if $H$ is considered to be a linear coupling function i.e., simple diffusion (say), then the above system \eqref{eq:1} may not be able to produce chimera states, as observed in locally coupled 2D grid of oscillators \cite{2d2}. In absence of coupling strength, individual SL systems exhibit limit cycle behaviour near the Hopf bifurcation point with intrinsic frequency $\alpha$.
	 
	 \par In order to analyze system \eqref{eq:1}, we first numerically show the existence of chimera states in the network of SL oscillators. To integrate the coupled system (\ref{eq:1}), we use Euler integration scheme with time step $\Delta t=0.01$. The initial conditions for each oscillator are taken as $x_{i,j,k}(0)=0.005(N-(i+j+k)),y_{i,j,k}(0)=0.008(N-(i+j+k))$ for $i,j,k=1,2,...,N$ along with some additional small random fluctuations. Here we considered $N=40$ in linear dimension of the 3D cubic lattice which accounts for $N^3=64000$ as the total number of oscillators in the entire 3D network. Figure \ref{isrevisefig1}(a) shows the snapshot of the state variables $x_{i,j,k}$ over the entire 3D lattice. Coexistence of coherence and incoherence and consequently the chimera pattern is easily identifiable from the figure. Snapshot of the state variables in 2D projection on the plane $k=40$ is depicted in Fig. \ref{isrevisefig1}(b), whereas the 1D snapshot (with $j=20, k=40$) of the oscillators distinguishing the domain of coherent and incoherent populations is shown in Fig. \ref{isrevisefig1}(c). The parameter values are fixed as $\alpha=1.0, \beta=-1.5, \tilde a=1.02, \epsilon=0.09$.
	
	\begin{figure*}[ht]
		\centerline{
			\includegraphics[scale=0.6]{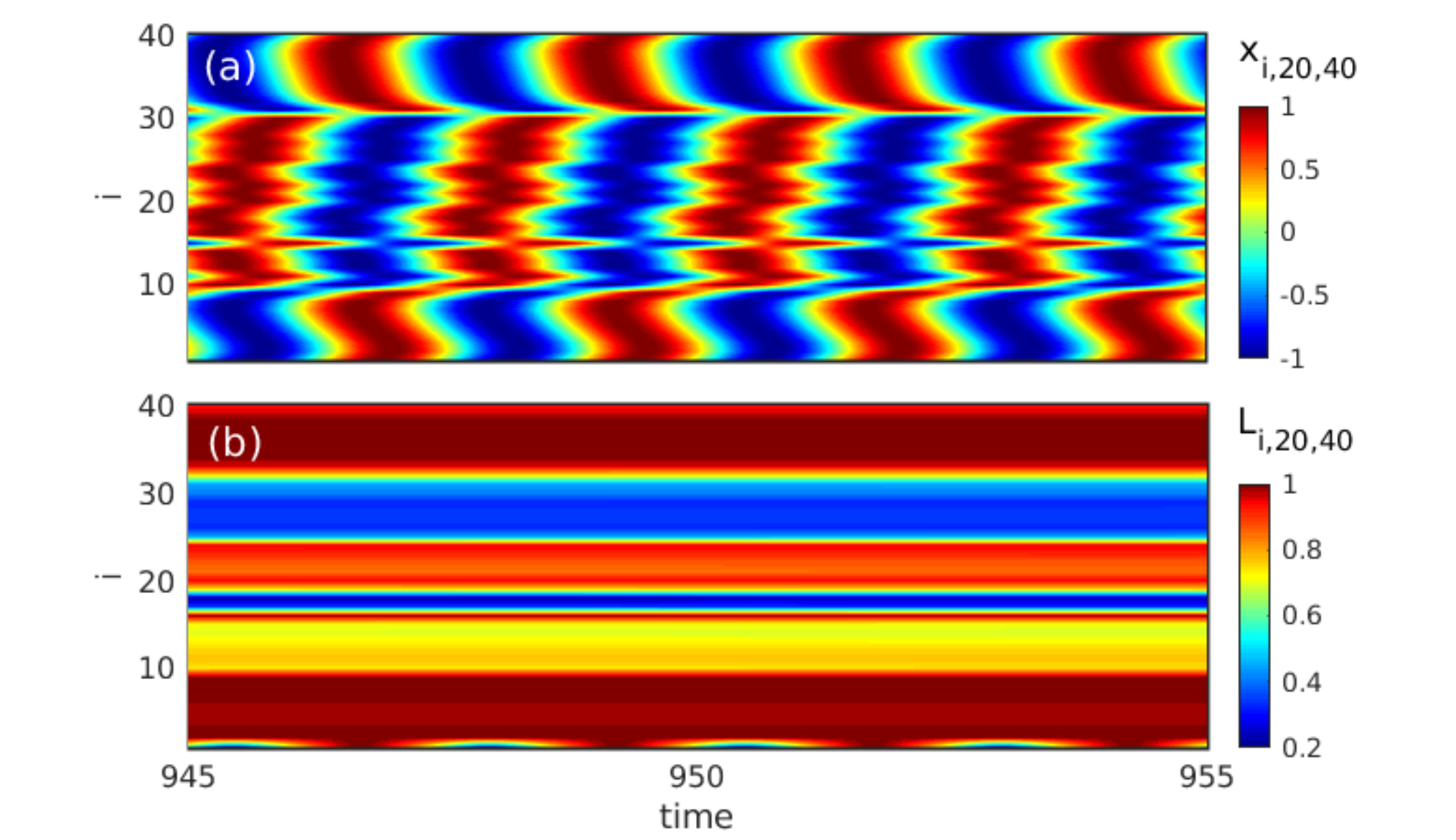}}
		\caption{ Spatiotemporal evaluation of (a) the state variable $x_{i,j,k}$, and (b) corresponding local order parameter $L_{i,j,k}$ of the SL oscillators. The plots are obtained along the cross-section $j=20,k=40$, keeping the coupling strength $\epsilon=0.09$ fixed in the chimera region. }
		\label{isrevisefig2}
	\end{figure*}

Figure \ref{isrevisefig2}(a) shows the spatiotemporal evolution of the state variable $x_{i,j,k} = Re(z_{i,j,k})$ for the oscillators situated along $j=20$ and $k=40$, while keeping the interaction strength as $\epsilon = 0.09$ fixed in the chimera region. The appearance of stationarity in the observed chimera pattern of 3D network of locally coupled SL oscillators is clearly visible from the figure. Moreover, to characterize the ordination of coherence and incoherence among the oscillators, we use the measure of local order parameter \cite{hr_bera3} $L_{i,j,k}$, which quantifies the degree of local ordering of the oscillators. The quantity is calculated using the usual formula
\begin{equation}
\begin{array}{lcl}
L_{i,j',k'} = \Big\lvert \frac{1}{2\delta}\sum \limits_{|i-i'|\leq \delta} e^{\hat{i}\Phi_{i'}} \Big\rvert,~~~~ j' = 20, k'=40.
\end{array}\label{eq: 4}
\end{equation} 
Here $\Phi_i = arctan(y_{i,j',k'}/x_{i,j',k'})$ is the geometric phase of the $(i,j',k')$-th oscillator, $\delta$ is the number of oscillators in the neighborhood of $(i,j',k')$-th oscillator and $\hat{i}=\sqrt{-1}$. In Fig. \ref{isrevisefig2}(b), the spatiotemporal evolution of the local order parameter $L_{i,j',k'}$ has been depicted. The value of the order parameter $L_{i,j',k'} \simeq 1$  signifies that the $(i,j',k')$-th oscillator belongs to the coherent part of the chimera state, whereas $L_{i,j',k'} < 1$ corresponds to incoherent parts, which is easily discernible from the figure. 

	\subsection{Analytical results: Ott-Antonsen Approach}\label{oa}
	
	To support the results obtained from our numerical experiments and to validate them in the continuum limit we go through the approach of Ott and Antonsen \cite{ott1,ott2}. Following the procedure discussed in \cite{2d2}, we can directly obtain the phase reduced model for the 3D coupled SL oscillators as 
	\begin{widetext}
		
		\begin{equation}\label{eq:6}
		\dot{\theta}_{i,j,k} = \omega'_{i,j,k} - \lambda \sum_{l=1}^{N}\sum_{m=1}^{N}\sum_{n=1}^{N} A_{ijklmn} \sin(\theta_{i,j,k} - \theta_{l,m,n} + \gamma),
		\end{equation}
		where $\omega'_{i,j,k}$ = $\omega_{i,j,k} + \epsilon\beta(\tilde{a}^2-1)$, $\lambda$ = $\frac{\epsilon}{6} (\tilde{a}^2 -1) \sqrt{1+\beta^2}$ and $\gamma = \tan^{-1}\beta$.  Since all the systems are identical here we consider $\omega_{i,j,k} = \omega, \forall i,j,k = 1,2, ..., N$ with $\omega = \alpha - \beta$ as in \cite{2d2}.
			\begin{figure*}[ht]
				\centerline{
					\includegraphics[scale=0.6]{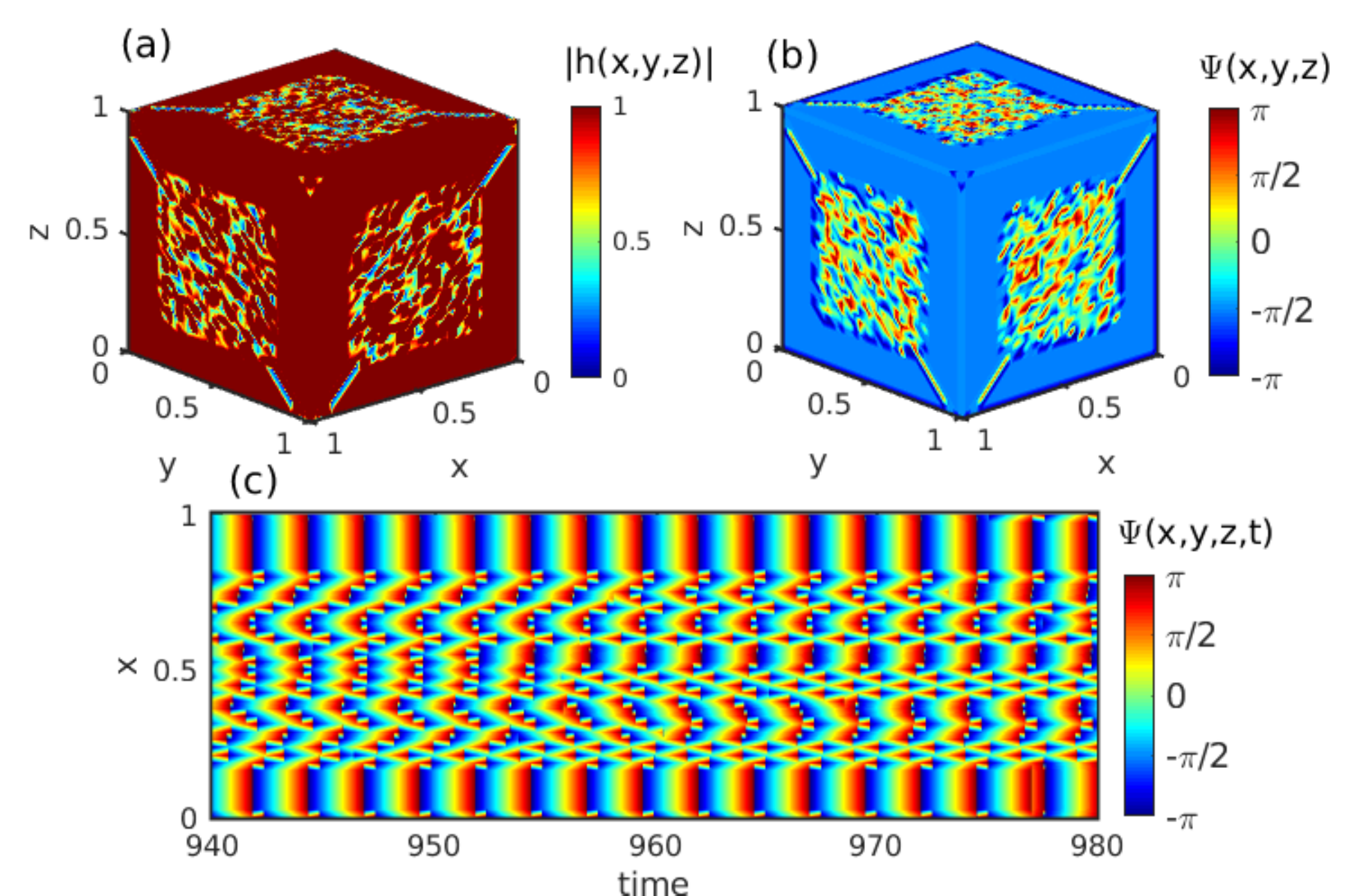}}
				\caption{ (a) Snapshots of the local order parameter $|h(x,y,z)|$ and (b) the local phase $\Psi(x,y,z)$ at a particular instant of time. (c) Spatiotemporal evolution of the phase $\Psi(x,y,z,t)$ for a long time interval, indicating stationarity of the observed chimera pattern.}
				\label{isrevisefig3}
			\end{figure*}
		It should be noted that the above equation can be rewritten in the continuous form as
		\begin{equation}
		\begin{array}{lcl}
		\frac{\partial\theta(x,y,z,t)}{\partial t} = \omega'- \lambda \int_{0}^{1}\int_{0}^{1}\int_{0}^{1} G(x-x',y-y',z-z') \sin(\theta(x,y,z,t) - \theta(x',y',z',t) + \gamma) dx' dy' dz'.
		\end{array}
		\end{equation}
		Here the coupling kernel $G$ can be written as $G(x-x',y-y',z-z') = H[\cos(\sqrt{(x-x')^2 + (y-y')^2 + (z-z')^2})2\pi - \cos(2\pi/N)]$ and $H(.)$ is the Heaviside step function.
		
		Considering the limiting case of an infinite number of oscillators, i.e. $N\rightarrow \infty$, the dynamics of the oscillators at time $t$ can be described by a probability density function $f(x,y,z,\theta,t)$ which satisfies the continuity equation, 
		\begin{equation}\label{eq: 8}
		\frac{\partial f}{\partial t} + \frac{\partial}{\partial \theta}(fv) = 0,
		\end{equation}
		where 
		\begin{equation}
		v = \frac{d\theta}{dt} = \omega' - \frac{1}{2\hat{i}}[re^{\hat{i}\theta} + \bar{r}e^{-\hat{i}\theta}].
		\end{equation}
		
		It is to be noted that here $r$ is the order parameter given by, 
		\begin{equation}
		\begin{array}{lcl}
		r(x,y,z,t) = \lambda e^{\hat{i}\gamma} \int_{0}^{1}\int_{0}^{1}\int_{0}^{1} G(x-x',y-y',z-z')\int_{0}^{2\pi} e^{-\hat{i}\theta}f(x',y',z',\theta,t)d\theta dx' dy' dz'.
		\end{array}	
		\end{equation}
		
		Then the probability density function $f(x,y,z,\theta,t)$ can be expanded in terms of the Fourier series taking into account the OA ansatz $f_n(x,y,z,\theta,t)$ = ${h(x,y,z,t)}^n$ as
		\begin{equation}\label{eq: 11}
		\begin{array}{lcl}
		f(x,y,z,\theta,t) = \frac{1}{2\pi}\left(1 + \sum_{n=1}^{\infty} {h(x,y,z,t)}^n e^{\hat{i}n\theta} + c.c.\right)\\\\~~~~~~~~~~~~~~
		= \frac{1}{2\pi}\left(1 + \sum_{n=1}^{\infty} (h^n e^{\hat{i}n\theta} + \bar{h}^n e^{-\hat{i}n\theta})\right).
		\end{array}
		\end{equation}
		
		Substituting \eqref{eq: 11} into \eqref{eq: 8}, we obtain the governing equation for the considered 3D network on the basis of OA reduction 
		\begin{equation}\label{eq: 12}
		\begin{array}{lcl}
		\frac{\partial h}{\partial t}=-\hat{i}\omega'h + \frac{1}{2}\left(\bar{r}h^2 + r\right),
		\end{array}
		\end{equation}
		
		where
		\begin{equation}\label{eq: 13}
		\begin{array}{lcl}
		r(x,y,z,t) = \lambda e^{\hat{i}\gamma} \int_{0}^{1}\int_{0}^{1}\int_{0}^{1} G(x-x',y-y',z-z') h(x',y',z',t) dx'dy' dz'.
		\end{array}	
		\end{equation}
		
\end{widetext}

The modulus and argument of the complex OA ansatz $h(x,y,z,t) = |h(x,y,z,t)|e^{-\hat{i}\Psi(x,y,z,t)}$ respectively provide information about the local order parameter and the local phase of the oscillators in the continuum limit.  The existence of the appropriate complex OA ansatz function $h(x,y,z,t)$ is the condition for the stability of the obtained chimera state.
\par Using Eqs. \eqref{eq: 12} and \eqref{eq: 13}, we have enlightened the appearance of chimera states through the complex OA reduction approach in a 3D network consisting of infinite number of locally coupled SL oscillators. Figure \ref{isrevisefig3} presents the results obtained through numerical integration of Eqs. \eqref{eq: 12} and \eqref{eq: 13}. In Figs. \ref{isrevisefig3}(a) and \ref{isrevisefig3}(b) respectively, we have plotted the snapshots of the local order parameter $|h(x,y,z)|$ and the local phase $\Psi(x,y,z)$ of the oscillators located in the considered 3D grid at a particular instant of time. Here in Fig. \ref{isrevisefig3}(a), $|h(x,y,z)|=1$ characterizes the domain of the phase locked oscillators (i.e., the coherent group), whereas, $|h(x,y,z)|<1$ characterizes the domain of oscillators with sparsely distributed phases (i.e., the incoherent group). Also from Fig. \ref{isrevisefig3}(b), it is evident that the coherent population has the same phase value, while the incoherent population has randomly distributed phases. Furthermore, in Fig. \ref{isrevisefig3}(c), we have plotted the space-time evolution of the phases $\Psi(x,y,z)$ to verify that the observed chimera patterns are stationary in time even in the continuum limit case. From these results, we can conclude that the OA reduction technique on the basis of infinite number of oscillators certainly  predicts the same dynamics, which was previously realized through numerical analysis.	

\section{Coupled Hindmarsh-Rose neuron model}\label{hrg}

Next we consider the dynamics of the network of a 3D cubic lattice of coupled Hindmarsh-Rose neuron model, consisting of $N^3=27000$ neuronal oscillators. The local dynamics is modeled through Hindmarsh-Rose (HR) \cite{hr_model} neuron, $F(X)=[ax^2-x^3-y-z, (a+\alpha)x^2-y, c(bx-z+e)]^T$ with coupling matrix $K=diag(\frac{\epsilon}{6},0,0)$ with coupling function $H(x', x)=(v_s-x')\Gamma(x)$, where $\Gamma(x)$ is the chemical synaptic function. After local coupling, the complete set of equations becomes,
\begin{widetext}
	\begin{equation}\label{hr}
	\begin{array}{lcl}
	\dot x_{i,j,k}=ax_{i,j,k}^{2}-x_{i,j,k}^{3}-y_{i,j,k}-z_{i,j,k}+\frac{\epsilon}{6}(v_{s}-x_{i,j,k})\{\Gamma(x_{i-1,j,k})+\Gamma(x_{i+1,j,k})\\~~~~~~~~+\Gamma(x_{i,j-1,k})+\Gamma(x_{i,j+1,k})+\Gamma(x_{i,j,k-1})+\Gamma(x_{i,j,k+1})\},\\
	\dot y_{i,j,k}=(a+\alpha)x_{i,j,k}^2-y_{i,j,k},\\
	\dot z_{i,j,k}=c(bx_{i,j,k}-z_{i,j,k}+e),~~~~i,j,k=1,2,...,N.
	\end{array}
	\end{equation} 
\end{widetext}
\begin{figure*}[ht]
	\centerline{
		\includegraphics[scale=0.55]{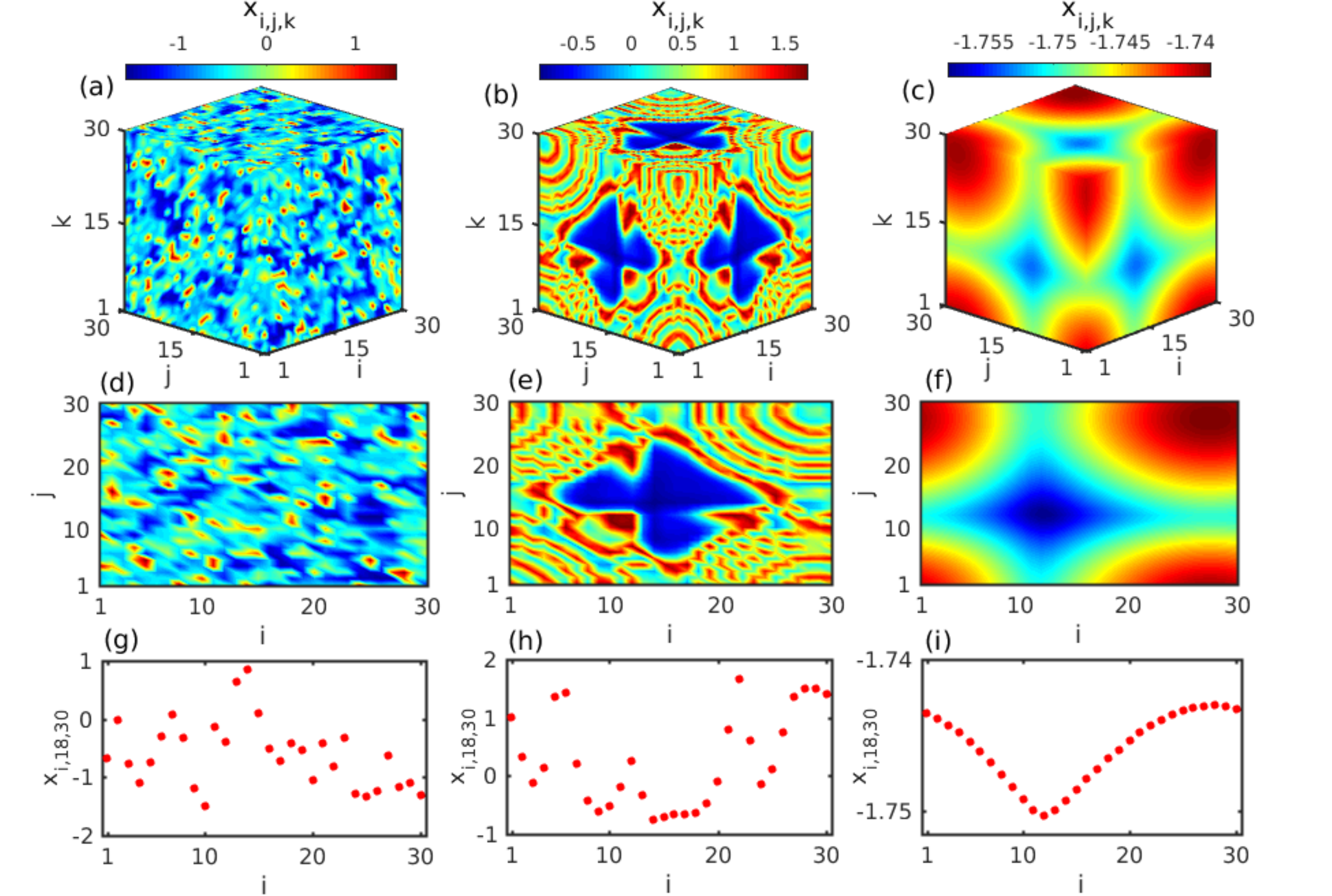}}
	\caption{Left, middle and right columns respectively represent (a, d, g) incoherent, (b, e, h) chimera and (c, f, i) coherent states emerging in a 3D cubic lattice of HR neuron. Top, middle and bottom row respectively represent the (a, b, c) 3D snapshots plotted in $(i, j, k)$ space, (d, e, f) 2D snapshots plotted in $(i, j)$ plane keeping $k = 30$ fixed, and (g, h, i) 1D snapshots plotted with respect to $i$ keeping $j = 18, k = 30$ fixed. The coupling strengths for incoherent, chimera and coherent states are respectively $\epsilon = 0.12, 1.2$ and $1.44$. All the snapshots are taken at time $t = 1495$.}
	\label{fig2}
\end{figure*}
The variable $x_{j,j,k}$ denotes the membrane potential of the $(i,j,k)$-th position of neuron in the 3D grid of neuronal network. The other two state variables $y_{i,j,k}$ and $z_{i,j,k}$ are associated with the ion transportation across the membrane potential in which  $y_{i,j,k}$ represents the fast current associated with $Na^+$ or $K^+$ ions and $z_{i,j,k}$ is the slow current associated with $Ca^+$ ions respectively. The parameter $c$ is controlled by the ratio of slow-fast dynamics. For the excitatory interaction, we choose $v_s=2$ so that the condition $x_{i,j,k}(t)<v_s$ is always maintained for all time $t$. The interaction function between the neurons is taken as a chemical synaptic function $\Gamma(x)$ which is a nonlinear sigmoidal input-output function, described by $\Gamma(x)=\frac{1}{1+e^{-\lambda(x-\Theta_s)}}$. The parameters $\lambda$ and $\Theta_s$ represent the slope of the sigmoidal function and synaptic firing threshold, respectively. The constant $\epsilon$ denotes the chemical synaptic interaction strength which summarizes how the information is delivered through the interaction among the neurons.  In particular, if electrical synapses are employed to interact between the neurons in 3D grid then the desired chimera states cannot be produced as verified in \cite{2d2} for 2D grid. We fix the local system parameter values as $a=2.8, \alpha=1.6, b=9.0, c=0.001, e=5.0$ in which the uncoupled neuron exhibits a the square wave bursting dynamics. We also choose the other parameters in the coupling function as $v_s=2.0, \lambda=10.0, \Theta_s=-0.25$. Next we study the behavior of all the neurons in the network by varying the coupling strength $\epsilon$ with $N=30$.

\par  Now we study the emergence of chimera states in a locally connected 3D grid consisting of $N=30$ HR oscillators in each direction numerically. So the total number of neurons corresponding to the linear dimension in the 3D grid formation is $N^3=27000$. For the numerical simulation, we again used the Euler integration scheme with time-step size $\Delta t=0.01$. The initial conditions for each node of the large neuronal network is chosen as $x_{i,j,k}(0)=0.001(N-(i+j+k)),y_{i,j,k}(0)=0.002(N-(i+j+k)),z_{i,j,k}(0)=0.003(N-(i+j+k))$ for $i,j,k=1,2,...,N$ with small added random fluctuations. Different dynamical states are portrayed in Fig. \ref{fig2} for various coupling strengths and the color bar denotes the values of the membrane potential $x_{i,j,k}$ attained by each neuron at a particular time $t=1495$. The left column of Fig. \ref{fig2} represents the incoherent states on 3D HR network for the coupling strength $\epsilon=0.12$. Figs. \ref{fig2}(a, d, g) represent the snapshots of the membrane potential $x_{i, j, k}$ in 3D space, 2D plane (taking cross-section along $k=30$) and 1D array (keeping $j=18,k=30$ fixed) respectively. In these figures, all the neurons  are randomly distributed which is a clear significance of incoherent states.  Similarly, the snapshots in the middle column of Fig. \ref{fig2} show the appearance of chimera states in different dimensions for the synaptic coupling strength $\epsilon=1.2$. Here the group of synchronized neurons are surrounded by the population of desynchronized neurons. In Figs. \ref{fig2}(b) and \ref{fig2}(e), the blue regions correspond to the coherent population of neurons and surrounding regions with mixed colors indicate the incoherent group of neurons. This feature is also evident in the 1D snapshot (Fig. \ref{fig2}(h)), where a group of neurons ($14\le i\le20$) follows a smooth profile while the rest are randomly scattered. This dynamical behavior is a strong indication of the existence of chimera patterns. Now for sufficiently higher values of the synaptic interaction strength, $\epsilon=1.44$, all the neurons in the coupled 3D network follow a coherent motion and the corresponding snapshots are shown in the right column of Figs. \ref{fig2} considering 3D, 2D and 1D cases. The continuous color variations (Figs. \ref{fig2}(c, f)) and the smooth profile (Fig. \ref{fig2}(i)) of each neuron are  bearing a strong resemblance of coherent states in the 3D neuronal network. 

\begin{figure*}[ht]
	\centerline{
		\includegraphics[scale=0.6]{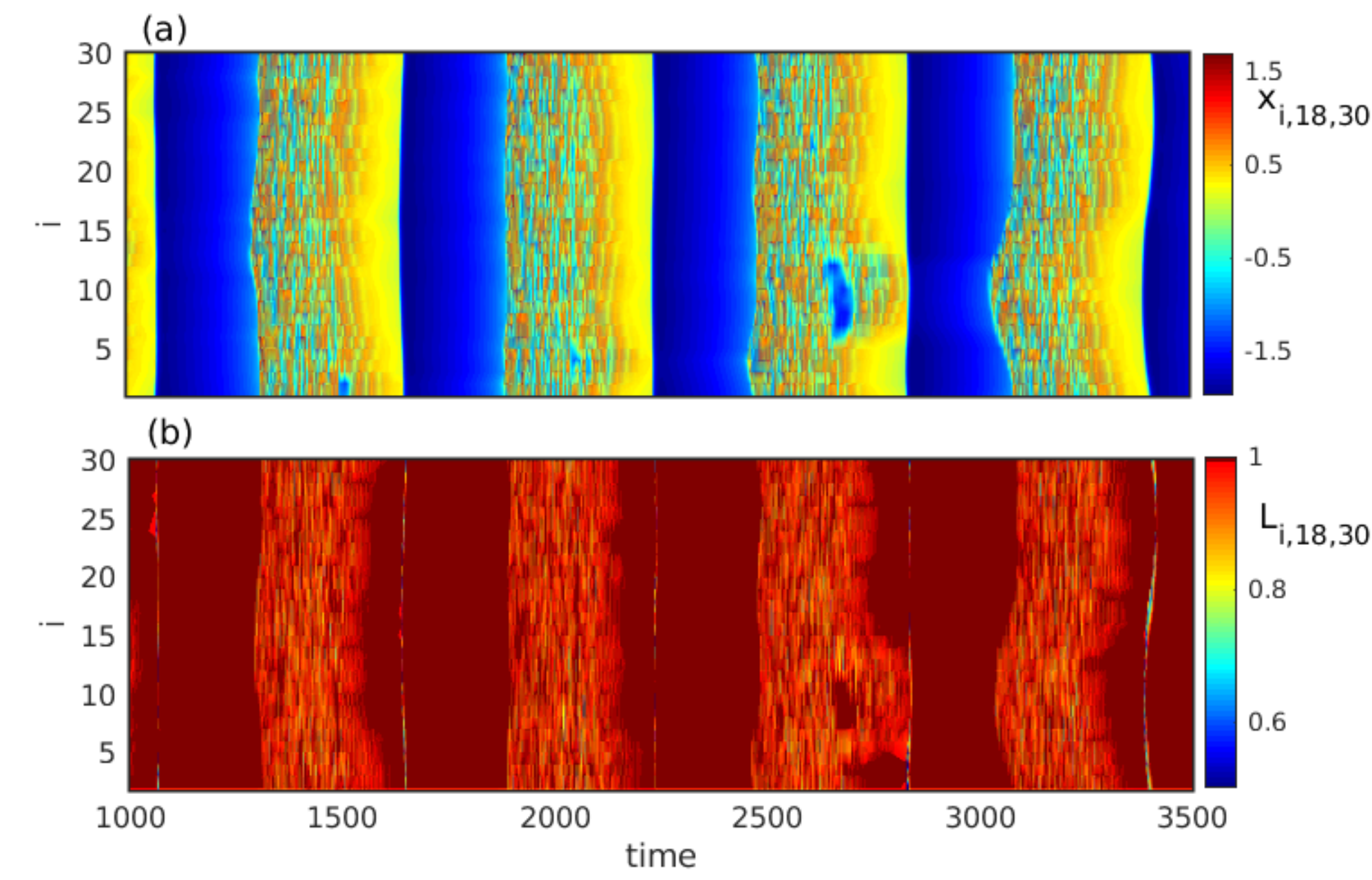}}
	\caption{Spatiotemporal evolution of (a) membrane potential $x_{i,j,k}$, and (b) corresponding local order parameter $L_{i,j,k}$ of the oscillators. The plots are obtained along the cross-section $j=18,k=30$, keeping the coupling strength $\epsilon=0.9$ fixed in the chimera region. }
	\label{fig3}
\end{figure*} 

\par Next we move on to investigate the stationarity of the chimera patterns and we find that our observed chimera state is not static, rather it varies in time. To prove the emergence of non-stationary chimera patterns, we plot the spatiotemporal dynamics of the neuronal network in Fig. \ref{fig3}(a) and its existence is characterized through the instantaneous order parameter plot (using Eq. \eqref{eq: 4} with $j'=18, k'=30$) in Fig. \ref{fig3}(b). Taking the cross-section along $j=18$ and $k=30$, the space-time plot is drawn at coupling strength $\epsilon=0.9$. In the chimera pattern, the blue region is the coherent state and mixed colors correspond to the incoherent states in Fig \ref{fig3}(a). These two populations oscillate with respect to time and the  instantaneous local order parameter signifies the appearance of this feature since it takes the value 1 from color bar (dark red region of Fig. \ref{fig3}(b)) for the coherent states and takes other lower values for the incoherent states.          

\begin{figure*}[ht]
	\centerline{
		\includegraphics[scale=0.58]{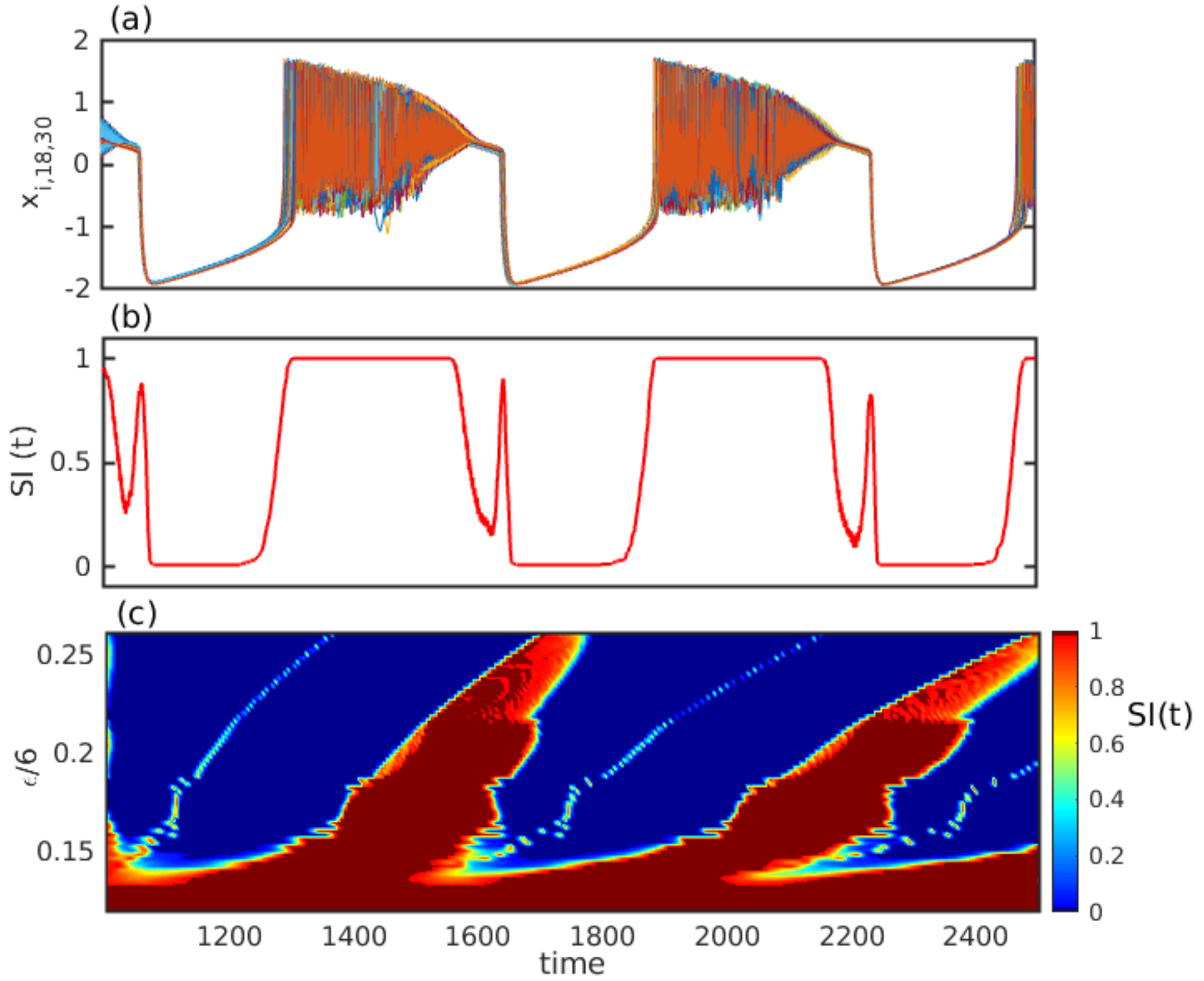}}
	\caption{(a) Time series of state variables $x$ of $N$ neurons situated on the 3D lattice along the cross-section $j=18,k=30$, (b) variation of $SI$ with respect to time of all the oscillators in the network for a particular coupling strength $\epsilon=0.9$. (c) Scenario of the strength of incoherence (SI) measure with respect to time and coupling strength.}
	\label{fig4}
\end{figure*}

To characterize the appearance of chimera states in 3D locally coupled network, we have considered the quantitative statistical measure based on the concept of strength of incoherence (SI) as developed by Gopal {\it et al.} \cite{lakshman_measure}. For this we begin by defining the transformations of the state variables $x_{i,j,k}$ to new variables,
\begin{widetext}
	\begin{equation}
	\begin{array}{lcl}
	w_{i,j,k} = \sqrt{(x_{i,j,k}-x_{i+1,j,k})^2 + (x_{i,j,k}-x_{i,j+1,k})^2 + (x_{i,j,k}-x_{i,j,k+1})^2},~~ i,j,k = 1, 2, ..., N.
	\end{array}
	\end{equation}
\end{widetext}
Next we divide the whole three dimensional medium into $M(=M_h \times M_w \times M_b)$ number of cubic bins each of equal volume $n_1(=N/M_h \times N/M_w \times N/M_b)$ and then calculate the local standard deviation $\sigma(p,q,r)$ in each of these bins as
\begin{widetext}
	\begin{equation}
	\begin{array}{lcl}
	\sigma(p,q,r)(t) = \sqrt{\frac{1}{n_1^3}\sum\limits_{i=n_1(p-1)+1}^{n_1p} ~~\sum\limits_{j=n_1(q-1)+1}^{n_1q}~~ \sum\limits_{k=n_1(r-1)+1}^{n_1r}(w_{i,j,k}-\langle w \rangle)^2},
	\end{array}
	\end{equation}
	where $\langle w \rangle = \frac{1}{N^3} \sum\limits_{i=1}^{N}\sum\limits_{j=1}^{N}\sum\limits_{k=1}^{N} w_{i,j,k}$, $p=1,2,...,M_h; q=1,2,...,M_w; r=1,2,...,M_b$. Then the strength of incoherence is defined as 
	
	\begin{equation}
	\begin{array}{lcl}
	SI(t) = 1-\frac{\sum\limits_{p=1}^{M_h}\sum\limits_{q=1}^{M_w}\sum\limits_{r=1}^{M_b}s(p,q,r)}{M}, ~~~~~s(p,q,r) = \Theta(\delta_1-\sigma(p,q,r)),
	\end{array}
	\end{equation}
\end{widetext}
where $\Theta(.)$ is the Heaviside step function and $\delta_1$ is the predefined threshold. Consequently, the values $SI(t)=1, SI(t)=0$ and $0<SI(t)<1$ characterize the incoherent, coherent and chimera states respectively.

%\end{widetext}
\par Now we study how the local dynamics of each of the neurons induce the non-stationary chimera patterns in the 3D lattice of neuronal network. The time evolution of the state variables $x_{i,j,k}$ corresponding to Fig. \ref{fig3}(a) is plotted in Fig. \ref{fig4}(a). From this figure it can be observed that, all the neurons are in a quiescent states while following the coherent motion (say, $1660<t<1840$), while for the incoherent state (say, $1840<t<2160$), all the neurons are in the spiking regime. Note that these spikes appear in a desynchronized manner, which help to induce the non-stationary chimera patterns. Through the instantaneous strength of incoherence $SI(t)$ measurement, the coherence-incoherence patterns are characterized. The variation of $SI(t)$ is shown in Fig. \ref{fig4}(b) corresponding to Fig. \ref{fig3}(a) and Fig. \ref{fig4}(a) for particular $\epsilon = 0.9$. Here $SI(t)$ takes the value $1$ for desynchronized states (spiking region) and takes the value $0$ for coherent states (quiescent states) while intermediate values refer to the chimera states. To reveal the coupling effect in the emergence of chimera patterns, we draw a phase diagram using $SI(t)$ measurement in Fig. \ref{fig4}(c) for varying $\epsilon$ by considering a long time interval. The color bar of Fig. \ref{fig4}(c) shows the variation of $SI(t)$ in which $SI(t)=1$ and $~0$ refer to the incoherent and coherent states respectively. The dark red and blue regions represent the incoherent and coherent states respectively while the intermediate colors correspond to the chimera states. For very low coupling values, all the neurons are in incoherent state for all times and chimera states (i.e., the co-existence of coherence and incoherence) appear after certain threshold value of $\frac{\epsilon}{6}>0.13$. Further increase in the coupling strength influences the dominating nature of the coherent states, leading to a shrinkage of incoherent and chimera regions.       
\begin{figure*}[ht]
	\centerline{
		\includegraphics[scale=0.6]{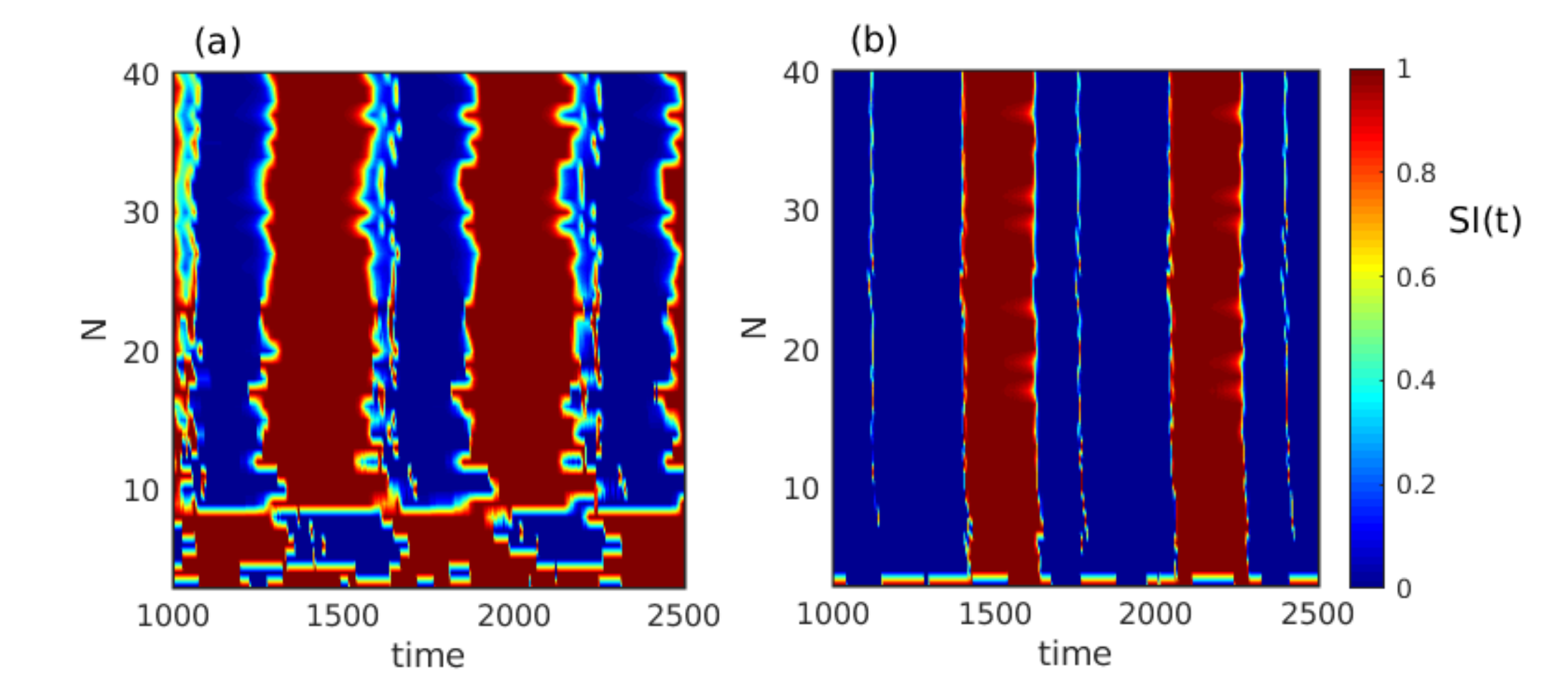}}
	\caption{Variation of $SI$ with respect to time obtained by varying the number of nodes in the network for two different coupling strengths (a) $\epsilon=0.9$ and (b) $\epsilon=1.08$.}
	\label{fig5}
\end{figure*} 

\par The effect of the network size on the emergence of non-stationary chimera patterns in the 3D neuronal network is also discussed in Fig. \ref{fig5}. For the coupling strengths $\epsilon=0.9$ and $1.08$, the dependency of $N$ on chimera pattern with respect to time is plotted in Fig. \ref{fig5}(a) and Fig. \ref{fig5}(b), respectively, and characterized through instantaneous strength of incoherence. From this figure it is clear that the imperfectness of non-stationary chimera patterns is reduced with an increase in the synaptic coupling strength and also the chimera patterns are independent of size after certain network size $N^3=9^3$.\\ 
\subsection{Synchronization threshold for 3D network model}\label{st}
Following the approach of Belykh {\it et al.} \cite{network_topo} to calculate the synchronization threshold in the case of chemical synaptic neuronal networks, here we can also calculate the threshold value in the case of our considered network with 3D lattice architecture. The important criterion to apply the mechanism proposed by Belykh {\it et al.} \cite{network_topo} for finding synchronization threshold  in any arbitrary network topology is that the number of signals received by each neuron should be equal to $n$ i.e., the degree of each of the neurons should be equal. Motivated by this, here we want to calculate the threshold for complete synchrony in case of our considered network (\ref{hr}) with 3D coupling topology, where the number of signals received by each neuron is equal to six, i.e. $n=6$. On the onset of complete synchronization each of the oscillators follow the single state $x_{i,j,k}(t) = x(t),~~ y_{i,j,k,}(t) = y(t),~~ z_{i,j,k}(t) = z(t)$ for  $i,j,k = 1, 2, ..., N$. We can say that the whole 3D network of $N^3$ oscillators  synchronizes, when each of the $N$ oscillators in each of the $i$th, $j$th and $k$th direction synchronizes. So for simplicity, we can calculate the threshold value by considering the oscillators along 1D, say along $i$ direction (for any arbitrary values of $j=j', k=k'$). Then if the $N$ number of oscillators along $(i,j',k')$ synchronizes, the whole network of $N^3$ oscillators synchronizes. We can rewrite the equation of the network model (\ref{hr}) along this specific $(i,j',k')$ direction as
\begin{widetext}
	\begin{equation}\label{hr1d}
	\begin{array}{lcl}
	\dot x_{i,j',k'}=ax_{i,j',k'}^{2}-x_{i,j',k'}^{3}-y_{i,j',k'}-z_{i,j',k'}+ \epsilon' (v_{s}-x_{i,j',k'})\sum\limits_{i'',j'',k'' = 1}^{N} C_{ii''j'j''k'k''}\Gamma(x_{i'',j'',k''}),\\
	\dot y_{i,j',k'}=(a+\alpha)x_{i,j',k'}^2-y_{i,j',k'},\\
	\dot z_{i,j',k'}=c(bx_{i,j',k'}-z_{i,j',k'}+e),~~~~i=1,2,...,N.
	\end{array}
	\end{equation} 
	Here $C_{ii''j'j''k'k''}$ is the connectivity matrix where $\sum\limits_{i'',j'',k'' = 1}^{N} C_{ii''j'j''k'k''} = n$ for each of the $(i,j',k')$-th oscillators and $\epsilon'=\frac{\epsilon}{6}$. 
	
	Adding and subtracting an additional term $\epsilon'(v_s-x_{i,j',k'})\sum\limits_{i'',j'',k'' = 1}^{N} C_{ii''j'j''k'k''} \Gamma(x_{i,j'',k''}) = n\epsilon'(v_s-x_{i,j',k'})  \Gamma(x_{i,j'',k''})$ to the $x$ equation of the above system, we get
	
	\begin{equation}\label{hr1d}
	\begin{array}{lcl}
	\dot x_{i,j',k'}=ax_{i,j',k'}^{2}-x_{i,j',k'}^{3}-y_{i,j',k'}-z_{i,j',k'} + n\epsilon'(v_{s}-x_{i,j',k'}) \Gamma(x_{i,j',k'}) \\~~~~~~~~~ + \epsilon' (v_{s}-x_{i,j',k'})\sum\limits_{i'',j'',k'' = 1}^{N} C_{ii''j'j''k'k''} (\Gamma(x_{i'',j'',k''})-\Gamma(x_{i,j',k'}))\\
	\dot y_{i,j',k'}=(a+\alpha)x_{i,j',k'}^2-y_{i,j',k'}\\
	\dot z_{i,j',k'}=c(bx_{i,j',k'}-z_{i,j',k'}+e),~~~~i=1,2,...,N.
	\end{array}
	\end{equation} 
\begin{figure*}[ht]
	\centerline{
		\includegraphics[scale=0.65]{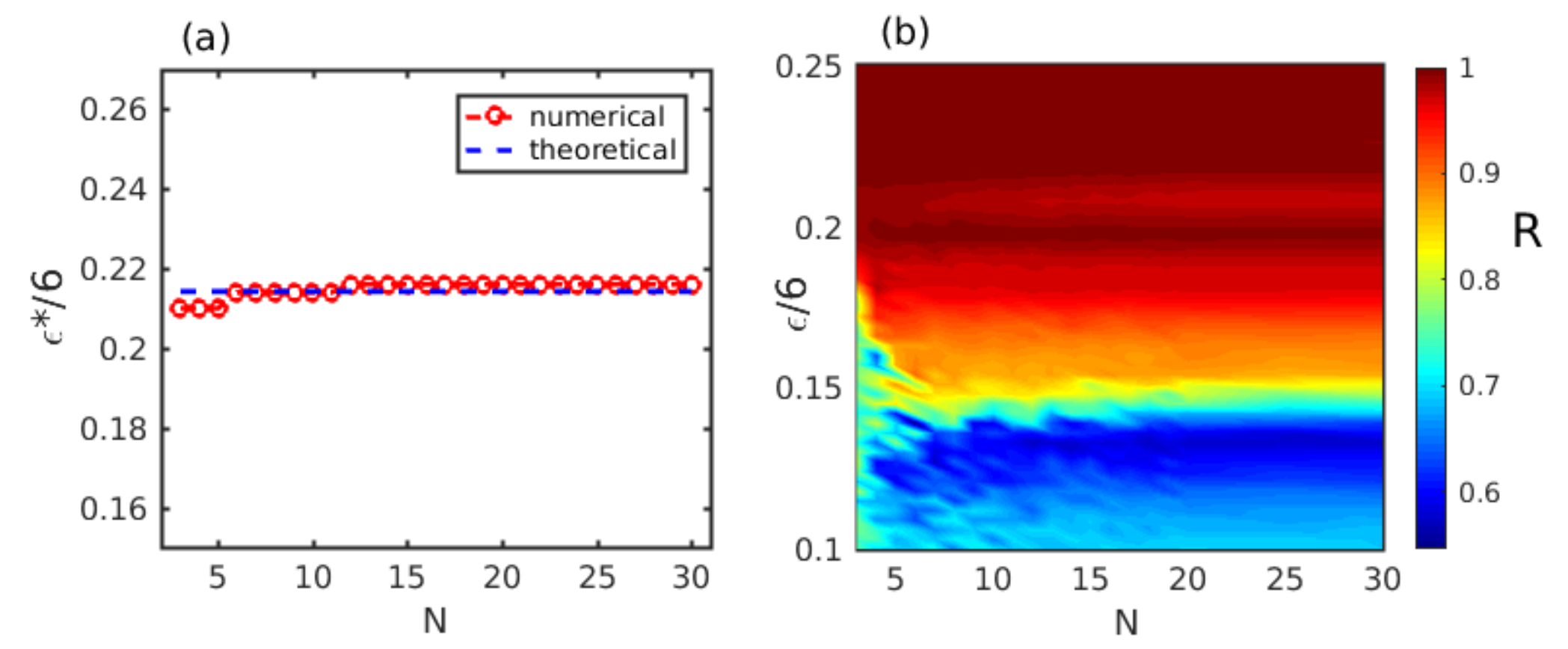}}
	\caption{(a) Synchronization threshold $\frac{\epsilon^*}{6}$ in 3D grid of network with respect to network size, (b) neuronal synchronization region in the phase diagram of $(N,\frac{\epsilon}{6})$ plane, characterized through global order parameter $R$. Here $N=30$ is the number of oscillators in one direction of the 3D lattice for which the network size is $N^3$.}
	\label{fig12}
\end{figure*} 	
	Introducing the differences between the neural oscillator coordinates $$X_{ii'',j'j'',k'k''} = x_{i'',j'',k''} - x_{i,j',k'}, Y_{ii'',j'j'',k'k''} = y_{i'',j'',k''} - y_{i,j',k'}, Z_{ii'',j'j'',k'k''} = z_{i'',j'',k''} - z_{i,j',k'},$$ we can derive the variational equations for transverse stability of the synchronization manifold as
	
	\begin{equation}\label{aa}
	\begin{array}{lcl}
	\dot{X}_{ii'',j'j'',k'k''} = (2ax-3x^2)X_{ii'',j'j'',k'k''} - Y_{ii'',j'j'',k'k''} - Z_{ii'',j'j'',k'k''} -n\epsilon'\Gamma(x) X_{ii'',j'j'',k'k''}\\~~~~~ + \epsilon' (v_s-x)\Gamma'(x)(nX_{ii'',j'j'',k'k''} + \sum\limits_{p',q',r'=1}^{N}\{C_{i''p'j''q'k''r'}X_{i''p',j''q',k''r'} - C_{ip'j'q'k'r'}X_{ip',j'q',k'r'} \}),\\
	\dot{Y}_{ii'',j'j'',k'k''} = 2(a+\alpha)xX_{ii'',j'j'',k'k''} - Y_{ii'',j'j'',k'k''},\\
	\dot{Z}_{ii'',j'j'',k'k''} = c(bX_{ii'',j'j'',k'k''} - Z_{ii'',j'j'',k'k''}).
	\end{array}
	\end{equation}
\end{widetext}
The origin $X_{ii'',j'j'',k'k''} = Y_{ii'',j'j'',k'k''} = Z_{ii'',j'j'',k'k''} = 0$ is an equilibrium point of the system (\ref{aa}) whose stability determines the stability of the synchronization manifold. The first coupling term $S_1 = -n\epsilon'\Gamma(x)X_{ii'',j'j'',k'k''}$ depends on the number of inputs $n$, whereas the second coupling term $S_2 = \epsilon'(v_s-x)\Gamma'(x)(.)$ depends on the coupling configuration. In terms of the original variable $x_{i,j',k'}$, the coupling matrix $\textbf{G} = \textbf{C} -n\textbf{I}$ is the negative of the Laplacian of the connected graph, which has one zero eigenvalue $\lambda_1$ and all other eigenvalues have non-positive real parts. Then we can write the stability equation on the synchronous manifold as

\begin{equation}\label{msf}
\begin{array}{lcl}
\dot{X} = (2ax-3x^2)X-Y-Z-\Omega(x)X,\\
\dot{Y} = 2(a+\alpha)xX-Y,\\
\dot{Z} = c(bX-Z),
\end{array}
\end{equation}
where $\Omega(x) = n\epsilon'\Gamma(x) - \epsilon'(v_s-x)\Gamma'(x)(n+\lambda_2)$, $\lambda_2$ being the eigenvalue with the largest real part. Equation \eqref{msf} is the same as obtained in \cite{network_topo} which allows us to directly write the estimate for synchronization threshold as $\epsilon'^* = \epsilon^*/n$, where $\epsilon^*=1.285$ is the synchronization threshold for two synaptically coupled HR neurons. This shows that the synchrony threshold does not depend on the lattice dimension of the interacting oscillators.

%\begin{widetext}

\par Now we have analytically obtained the bound for fully coherent region through the linear stability analysis. To verify it with our numerical results, we have plotted the synchronization threshold $\epsilon'^*$ against the network size $N$ in Fig. \ref{fig12}(a), where dashed blue line and open red circle lines represent the theoretical and numerical results respectively. To numerically calculate the measure of the coherence level in the 3D locally connected neurons, we use Kuramoto order parameter $R$, defined as

\begin{equation}
\begin{array}{lcl}
R = \langle R(t) \rangle_t = \langle \lvert \frac{1}{N^3}\sum\limits_{i=1}^N\sum\limits_{j=1}^N\sum\limits_{k=1}^N e^{\hat{i}\Phi_{i,j,k}(t)}\rvert \rangle_t,
\end{array}
\end{equation}
where $\Phi_{i,j,k}$ is the phase of the $(i,j,k)$-th neuron and $\hat{i}=\sqrt{-1}$. $R \simeq 1$ corresponds to fully synchronized motion, whereas smaller values of $R$ characterizes the desynchronized motion of the coupled network. In Fig. \ref{fig12}(a), the critical coupling strength $\epsilon'^*$, for which $R \simeq 1$, has been plotted and from the figure it is evident that the numerically obtained bounds are in good agreement with the analytically obtained ones.
The lines in the figure demarcate the  region of synchrony (coherence) and also show that the critical neuronal synchrony threshold is quite independent of the network size. Moreover, for better realization, we have plotted the values of $R$ in the ($N,\frac{\epsilon}{6}$) parameter space in Fig. \ref{fig12}(b).
 
\section{Conclusion}\label{con}
In this paper we have studied the existence and emergence of chimera patterns in three dimensional grids of symmetrically coupled identical oscillators. The nodes of the network are assumed to interact nonlinearly in the locally connected network. In the first part, we have shown the existence of stationary chimera states in the network of identically coupled Stuart-Landau oscillators. Using the OA approach, we have investigated the chimera patterns in the continuum limit corresponding to infinite number of oscillators and it is observed that the local order parameter and the local phase of the oscillators, obtained through the analysis of OA reduction technique, validate the numerical results. Then, by considering a realistic neuronal interacting medium as chemical synaptic coupling function, we enunciated the emergence of chimera patterns in the 3D grid of neuronal network. The mathematical form of the chemical synapse is of the sigmoidal type nonlinear function. Hindmarsh-Rose bursting neurons are considered as the local dynamics of the neuronal network. Here the important observation is the emergence of non-stationary chimera patterns. The combination of slow-fast dynamical features of each neuron and the nonlinear interaction function among them leads to the appearance of such chimera patterns. Using different characterizing quantities, such as strength of incoherence and local order parameter, we have characterized various collective dynamical states and articulated the transition scenarios among them.  We can say that the nonlinearity present in the interaction function acts as an essential condition for the existence of chimera states in our considered network. Further, we have also calculated the synchronization threshold for HR network interacting in 3D lattice architecture. Our finding reveals that the synchrony threshold is independent of the lattice dimension, which we consider to be an interesting observation. The central aim of the present work is to establish the generalization of the dynamics of one dimensional and two dimensional networks to three dimensional networks, truly inspired from the complex connectivity structure of the brain network.  Our findings may be generalized in more 3D complex structures which are related to  connectivity formation in the brain. The present study may give a better knowledge to understand the different neuronal disorder diseases. The effect of time-delay in the local coupling function with different 3D network topologies can be extended for further studies.  Also \textit{how does the increase in the lattice dimension affect the observed nonlinear phenomena} will be an interesting question to discuss in the near future.   
\medskip

%\clearpage

\par {\bf Acknowledgments:}
DG was supported by DST-SERB (Department of Science and Technology), Government of India (Project no. EMR/2016/001039). The work of ML is supported by a DST-SERB Distinguished Fellowship program.

\end{document}